\documentclass[12pt, draftclsnofoot, onecolumn]{IEEEtran}
\ifCLASSINFOpdf
\else
\fi
\usepackage{cite}
\usepackage{amsmath,amssymb,amsfonts}
\usepackage{algorithmic}
\usepackage{graphicx}
\usepackage{textcomp}
\usepackage{xcolor}
\usepackage{amssymb}
\usepackage{amsthm}
\usepackage{subcaption}
\usepackage{mathtools}

\begin{document}

%
\title{Performance Evaluation of Cooperative NOMA-based Improved Hybrid SWIPT Protocol}
%
%
%

\author{Ahmed Al Amin
        and~Soo Young Shin,~\IEEEmembership{Senior Member,~IEEE}
\thanks{The authors are with the WENS Laboratory, Department of IT Convergence Engineering, Kumoh National Institute of Technology, Gumi 39177, South Korea (e-mail: amin@kumoh.ac.kr; wdragon@kumoh.ac.kr).}
\thanks{Corresponding author: Soo Young Shin (wdragon@kumoh.ac.kr)}
\thanks{}}

%
%

\markboth{}%
{Amin \MakeLowercase{\textit{et al.}}: Bare Demo of IEEEtran.cls for IEEE Journals}
%



\maketitle

\begin{abstract}
This study proposes the integration of a cooperative non-orthogonal multiple access (CNOMA) and improved hybrid simultaneous wireless information and power transfer (IHS SWIPT) protocol (termed as CNOMA-IHS) to enhance the spectral efficiency (SE) of a downlink (DL) CNOMA communication system. CNOMA-IHS scheme can enhance the ergodic sum capacity (ESC) and energy efficiency (EE) of DL CNOMA by transferring additional symbols towards the users and energize the relay operation as well without any additional resources (e.g., time slot/frequency/code). The analytical and simulation results indicate that the proposed CNOMA-IHS scheme outperforms other existing SWIPT-based schemes (e.g., CNOMA with hybrid SWIPT, CNOMA with power-splitting SWIPT, wireless-powered CNOMA, CNOMA with time switching SWIPT, and orthogonal multiple access with IHS SWIPT) in terms of the ESC. Moreover, the CNOMA-IHS scheme also enhances EE compared with other conventional TS-SWIPT-based schemes, which is also illustrated by the simulation results. In addition, the proposed CNOMA-IHS scheme with the considered EE optimization technique outplayed the proposed CNOMA-IHS scheme without EE optimization and other existing TS-SWIPT-based schemes in terms of EE.   
\end{abstract}


%
\IEEEpeerreviewmaketitle

\section{Introduction}
%
%
%
%
\IEEEPARstart{N}{on-orthogonal} multiple access (NOMA) transmits simultaneous data to multiple users using power or code domain multiplexing technique without additional radio resources [1]. NOMA can provide high spectral efficiency to a considerable number of users [2]. There are two primary types of NOMA techniques: code-domain NOMA and power-domain (PD) NOMA. Code-domain NOMA facilitates user separation at the receiver end by introducing redundancies via coding or spreading. In contrast, PD NOMA can perform successive interference cancellation (SIC) for users having better channel conditions [3-5]. This ensures flexibility in resource allocation as well as improving the performance of NOMA [6]. Hence, PD-NOMA has been considered in this study, and, hereafter, NOMA refers to PD-NOMA in this study.
\par
In downlink (DL) NOMA, superposed signals are transmitted to users simultaneously [1-6]. After receiving the signals, the cell center user (CCU) applies SIC to decode the received signals [4-5]. According to NOMA, the signal power of the cell edge user (CEU) is always higher than that of a CCU. Hence, the signal is directly decoded at the CEU, and the signal of the CCU is considered as noise. NOMA can be applicable for future cellular communication applications [2-10]. One of the research fields related to NOMA is cooperative NOMA (CNOMA) [6-14]. CNOMA can be classified into two major categories. Strong users, such as CCUs, act as relays in the first category, and in the other category, NOMA users are assisted by dedicated relays [6]. In this study, the first category of CNOMA is considered because dedicated relays are not universally available. Moreover, the user-assisted CNOMA enhances the coverage area and data reliability of a wireless communication system considering users with the best channel conditions as relays [6-8]. In the case of the conventional user-assisted CNOMA, CCU has a better channel condition than CEU. Hence, CCU has been considered a relay to enhance the coverage area and data reliability of the CEU in case of user-assisted relaying [9]. The simultaneous wireless information power transfer (SWIPT) protocol can extract energy to perform energy harvesting (EH) from ambient radio frequency signals and also transfer information simultaneously [10-14]. 
\par
The major challenge of the CCU is to reduce the battery drainage issue while performing the relay operation [10-13]. Such drainage can cause CCU equipment to turn off and terminate the relay operation. Thus, the performance of the network is degrading significantly. To mitigate this problem, a hybrid SWIPT (HS) protocol was proposed and considered for CNOMA (CNOMA-HS) in [11-13]. Moreover, CNOMA-HS scheme can provide more harvested energy than conventional power splitting (PS) and time switching (TS) based SWIPT protocol because HS protocol is a combination of TS and PS based SWIPT protocol [11--13]. CNOMA with TS and PS-based SWIPT protocol was proposed in recent studies [14--15]. Furthermore, wireless powered CNOMA (WP-CNOMA) was proposed to empower the relay based CNOMA and also enhance the throughput by the proposed technique [14]. In addition, an optimization technique for outage probabilities and ESC in the case of CNOMA with SWIPT protocol was discussed in [15]. But the improvement of user channel capacities and ESC of the CNOMA-HS-SWIPT scheme were not discussed in [11--15]. The main challenges of the CNOMS-HS scheme are the degradation of user channel capacities and ESC because TS and PS SWIPT are combined in the DL CNOMA-HS-SWIPT scheme. Moreover, no suitable technique was proposed for the enhancement of ESC in the case of the DL CNOMA-HS scheme [11-13,15]. Therefore, a suitable HS SWIPT protocol is required which can provide sufficient EH for decode and forward (DF) relay operation and enhance user channel capacities, as well as the ESC of the DL CNOMA without any extra resources (Time/frequency/code) or interference issues [16-17]. Hence, suitable transmission strategies should be integrated with the DL CNOMA-HS scheme so that the channel capacities, along with the ESC, are improved significantly. \par
Energy efficiency (EE) is another vital factor in case of future wireless communication system in case of wireless information and power transfer [18]. Moreover, SWIPT protocol provides the possibility of improving the EE [19]. Thus, EE improvement is a vital issue in case of CNOMA with SWIPT protocol based scheme [18--19]. But in previous studies, the enhancement of user channel capacities, ESC, and EE of the CNOMA with SWIPT protocol has not been extensively explored. To address this issue, a suitable HS protocol and scheme are required to enhance the ESC and EE using the idle link without any extra resources or interference.    
\par

To enhance the user channel capacities, ESC, and EE of a hybrid SWIPT protocol with DL CNOMA, a novel improved hybrid SWIPT (IHS) protocol has been proposed in this study. Moreover, CNOMA is integrated with the IHS protocol which is termed as CNOMA-IHS scheme that enhances the ESC of DL CNOMA cellular networks. This scheme energizes the CCU for DF relay operations and transmits additional symbols to CCU and CEU, which enhance the channel capacity of CCU and CEU using different transmission strategies without consuming any additional resources or introduce any interference issue. This mainly improves the user channel capacities of the proposed scheme compared to the existing schemes. Consequently, the proposed scheme enhances the ESC in comparison with other conventional SWIPT protocol-based schemes (e.g., CNOMA with HS [11], CNOMA with power-splitting SWIPT (CNOMA-PS) [20], WP-CNOMA[14], CNOMA with time-switching SWIPT (CNOMA-TS) [20], and orthogonal multiple access with IHS (termed as OMA-IHS) schemes). Furthermore, the EE improvement of the proposed scheme is also evaluated and compared to other conventional SWIPT protocol-based schemes as well. As fraction of block time for EH is superior factor than the power splitting ratio in case of hybrid SWIPT protocol [11-13,15]. Hence, fraction of block time for EH based EE optimization technique is considered in this study [18]. In addition, the proposed CNOMA-IHS scheme with the considered EE optimization technique outplayed the proposed CNOMA-IHS scheme without EE optimization technique and other existing SWIPT-based schemes in terms of EE.

\par
 The primary contributions of this study are as follows: 
\begin{itemize}
\item In this study, CNOMA-IHS scheme is proposed considering the CCU as a relay. In addition, CNOMA-IHS scheme reduces the battery drainage issue of the CCU and improves the user capacities, ESC, and EE significantly.

\item The ESC of the proposed CNOMA-IHS scheme is analyzed and compared with the existing SWIPT based schemes (e.g., CNOMA-HS[11], CNOMA-PS[20], WP-CNOMA[14], CNOMA-TS[20], and OMA-IHS) as well. 

\item Moreover, the impact of the different parameters of the SWIPT protocol on the ESC of the proposed CNOMA-IHS scheme are also evaluated and compared with the existing SWIPT-based schemes [11,14,20]. 

\item Using analytical and simulations results, the ESC improvement of the proposed scheme compared with other existing schemes is explicitly evaluated [11,14,20].

\item EE of the proposed CNOMA-IHS scheme is evaluated for the proposed scheme and compared with existing SWIPT-based schemes (e.g., CNOMA-HS[11], WP-CNOMA[14], CNOMA-TS[20], and OMA-IHS). Moreover, the impact of fraction of block time for EH on the EE is also evaluated and compared with the existing schemes. Furthermore, the proposed CNOMA-IHS scheme with considered EE optimization is also compared with the proposed scheme without considered EE optimization and other existing SWIPT-based schemes in terms of EE.

\end{itemize}

The remainder of this paper is organized as follows: 
Section 2 describes the CNOMA-IHS scheme using the system model. Section 3 evaluates the result analysis. Section 4 concludes the paper.  

\par
\section{System Model and IHS Protocol}
A system model of the DL CNOMA-IHS scheme using a base station (BS) as a source ($S$) and two users (a CCU called $UE_1$ and a CEU called $UE_2$) in a single-cell scenario is considered. The user-assisted energy-constrained relay used to enhance the data reliability and coverage area of the network is denoted as $UE_1$ [11--13]. Furthermore, $UE_1$ conducts IHS-based EH to perform DF relaying for $UE_2$ by the harvested energy. $S$, $UE_1$, and $UE_2$ are considered as single antenna devices. The system model of the proposed scheme is illustrated in Figure  1. The subscripts $S$, $1$, and $2$ correspond to $S$, $UE_1$, and $UE_2$, respectively. Here, $d_{S,1}$ and $d_{S,2}$ denote the corresponding normalized distances of $UE_1$ and $UE_2$ from $S$, as depicted in Figure  1. Furthermore, $d_{1,2}$ denotes the normalized distance between $UE_1$ and $UE_2$. The independent Rayleigh fading channel coefficients corresponding to the $S$-to-$UE_1$, $S$-to-$UE_2$, and $UE_1$-to-$UE_2$ links are denoted by $h_{S,1}$, $h_{S,2}$, and $h_{1,2}$, respectively. The channel coefficient $h_{i,j} \sim CN(0,\lambda_{i,j})$ between any two nodes $i$ and $j$ $(i,j \epsilon \{S, UE_1, UE_2\})$ and $i \neq j$) is related to the Rayleigh fading channel, along with the Gaussian random noise with variance $\sigma^2$ and zero mean, which are considered in this study [11--13,20]. The path loss exponent of the proposed system model is represented by $v$, and the distance in meters is denoted by $d_{i,j}$. Moreover, $\lambda_{S,1} > \lambda_{S,2}$ and $\lambda_{1,2} > \lambda_{S,2}$ because $d_{S,1} < d_{S,2}$ and $d_{1,2} < d_{S,2}$ have been considered in this study. Furthermore, all the Rayleigh fading channel gains are considered complex channel coefficients [11--13,20]. Based on the principle of DL NOMA, $p_{N}$ and $p_F$ denote the powers allocated from $S$ to $UE_1$ and $UE_2$, respectively, where $p_F$ \textgreater $p_{N}$ because $d_{S,1} < d_{S,2}$, and $p_{N}+p_{F}=1$. In addition, $P$ denotes the total transmission power of $S$. $p_N$ and $p_F$ can be determined by following equations [11, 21]: 
\begin{equation}
p_N= \frac{2^{2R_{th,1}}-1}{2^{2R_{th,1}+2R_{th,2}}-1}, 
\end{equation} 
\begin{equation}
p_F= 1-p_N, 
\end{equation}
Where $R_{th,1}$ and $R_{th,2}$ are the targeted data rate of $UE_1$ and $UE_2$, respectively. 
Moreover, $\theta$ ($0<\theta<1$) and $\delta$ ($0<\delta<1$) are the fraction of block time for energy harvesting and power allocation factor for the proposed IHS protocol, respectively [11--13]. A symbol $x_1$ is transmitted to $UE_1$ for EH and is simultaneously transmitted to $UE_2$ for information transmission during phase-1 with $P$. The NOMA-based superimposed signals are transferred toward the users during phase-1 as well. So, $x_2$ and $x_3$ are transmitted as superimposed signal towards the users during phase-1. Furthermore, a power splitting (PS)-based EH is performed during phase-1 by $UE_1$ using $\delta$. Moreover, $x_2$ is decoded using the harvested energy ($1-\delta$) by $UE_1$. Furthermore, $x_3$ is directly decoded by $UE_2$ because $p_F$ \textgreater $p_{N}$ during phase 1. 
\begin{figure}[!h]
\centering
\includegraphics[width=0.6\textwidth]{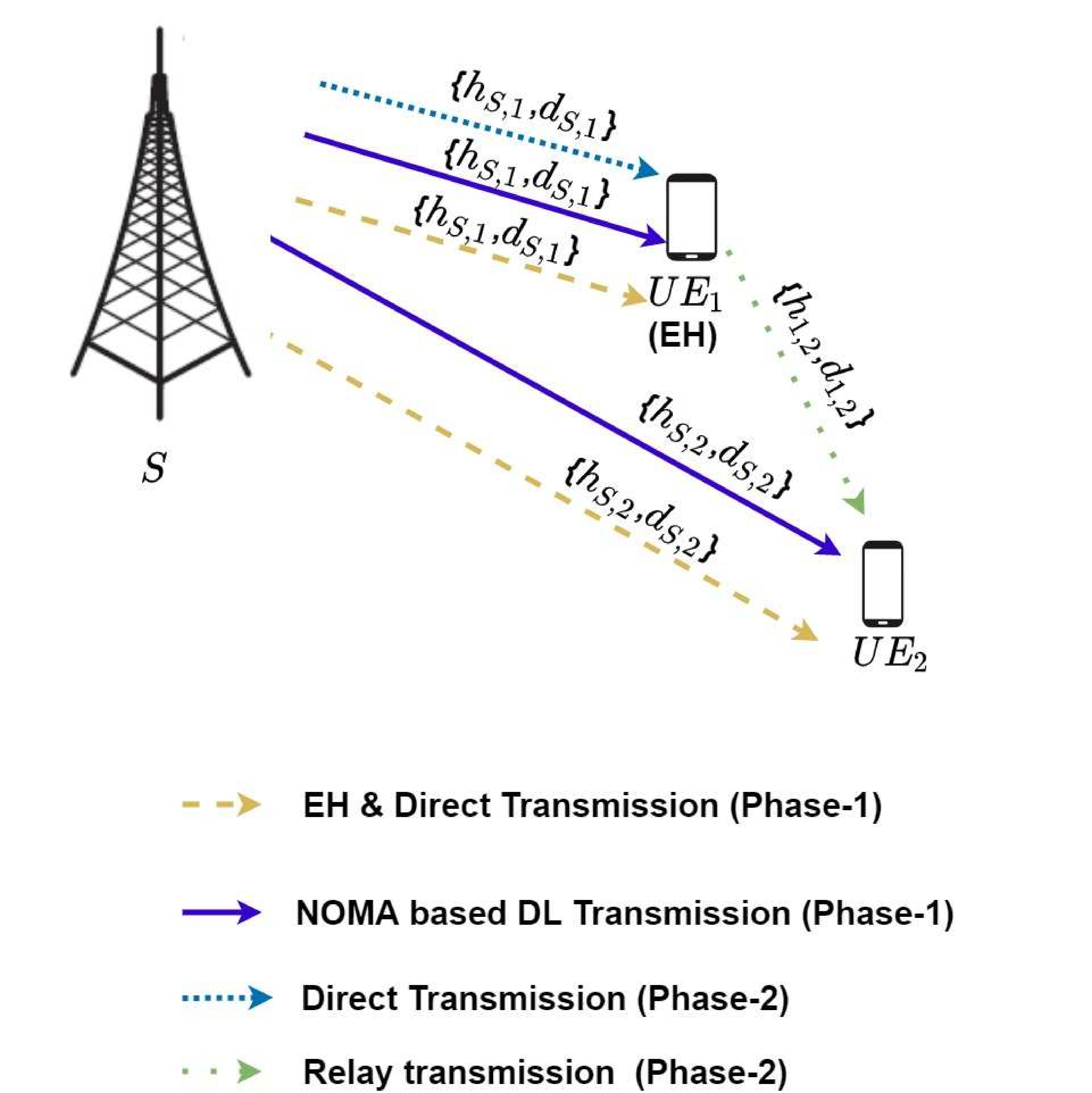}
\caption{System model of proposed CNOMA-IHS scheme.}
\label{image-myimage}
\end{figure}
\begin{figure}[!h]
\centering
\includegraphics[width=0.5\textwidth]{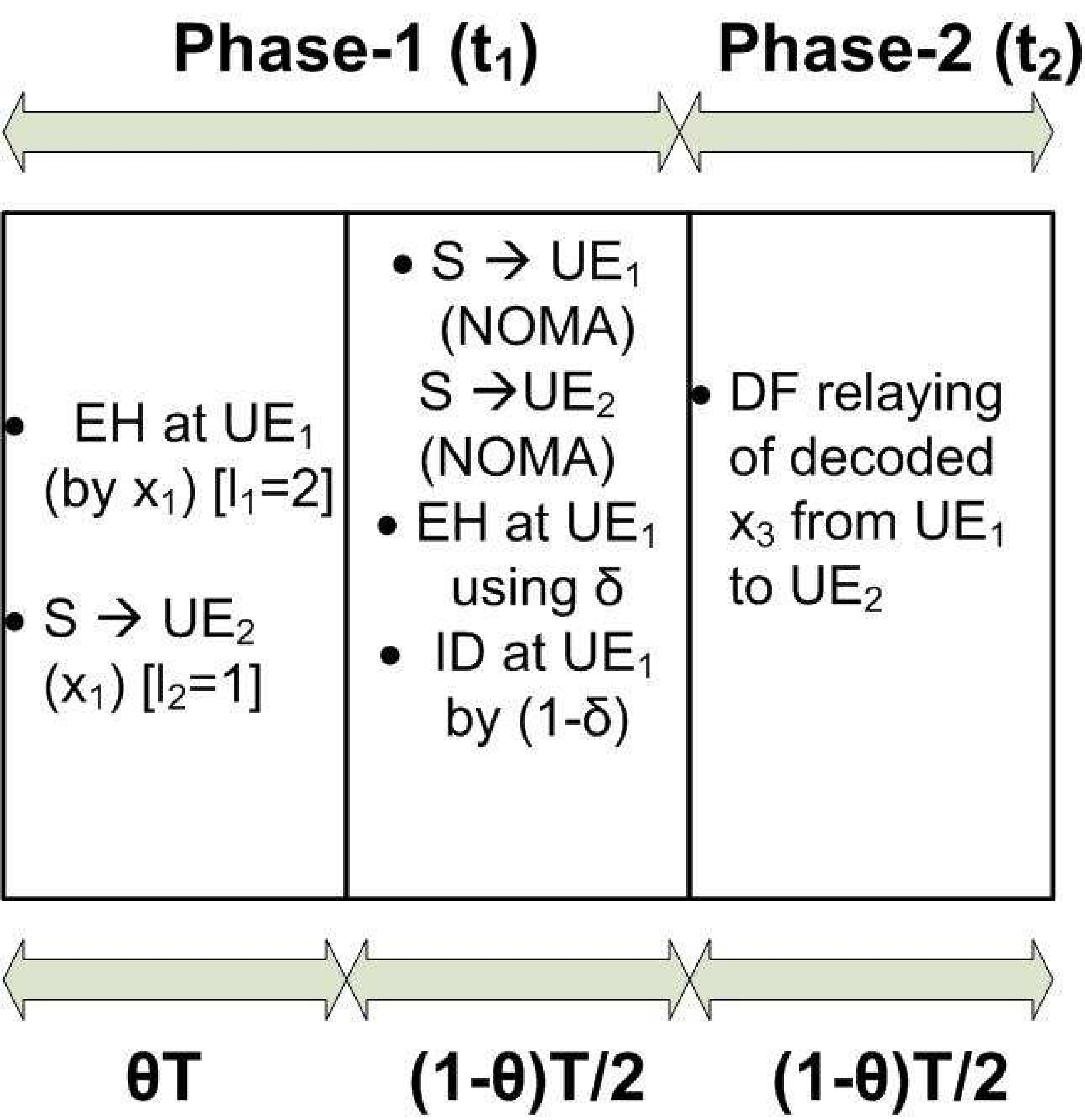}
\caption{Proposed protocol of CNOMA-IHS scheme.}
\label{image-myimage}
\end{figure}
\par
During phase 2, $UE_1$ performs the DF relaying of $x_3$ toward $UE_2$ to enhance data reliability and coverage area of the cellular network. In addition, an additional symbol $x_4$ by $p_N$ is transmitted to $UE_1$ from $S$ during second segment of phase 2. The transmission strategies are illustrated in Figure  1.

The proposed protocol of the CNOMA-IHS scheme is depicted in Figure 2, where $T$ denotes the total time duration required for a complete DL transmission. During the first segment of phase 1 ($\theta T$ duration), $x_1$ is transmitted to $UE_1$ and $UE_2$, simultaneously. Moreover, $x_2$ and $x_3$ are transmitted using the superimposed NOMA signal to $UE_1$ and $UE_2$ during the second segment of phase 1 (($(1-\theta)T/3$)). During first segment of phase 2 ($(1-\theta)T/3$), $UE_1$ relays decoded $x_3$ ($\hat x_3$) to $UE_2$ to improve data reliability and coverage area of the cellular network. In addition, $x_4$ is also transmitted from $S$ to $UE_1$ during second segment of phase 2 ($(1-\theta)T/3$). The detailed descriptions of the different phases along with the signal-to-interference plus noise ratio (SINR) equations for different symbols are given in the following subsections:
\subsection{Phase-1($t_1$)}
In the proposed CNOMA-IHS scheme, $x_1$ is transmitted from $S$ to the users using $A_1$ signal with transmitted power $P$. So, $A_1$ can be expressed as below:
\begin{equation}
A_1=\sqrt{P}x_1,
\end{equation}
where $x_1$ denotes a data symbol for $UE_2$. In addition, $x_1$ is transmitted to $UE_1$, which harvests the energy during the first segment of phase 1 ($\theta T$). Moreover, $A_1$ is received by $UE_2$ and $UE_2$ decodes $x_1$ during the first segment of phase 1 ($\theta T$). The received signals at $UE_1$ and $UE_2$ are expressed by the following equations: 
\begin{equation}
y_{S,1}^{t_1} = (\sqrt{P}x_1)h_{S,1} + n_1,
\end{equation}
\begin{equation}
y_{S,2}^{t_1} = (\sqrt{P}x_1)h_{S,2}+ n_2,
\end{equation}
where $n_1 \sim CN(0,\sigma^2)$ and $n_2 \sim CN(0,\sigma^2)$ are the complex white Gaussian noises at $UE_1$ and $UE_2$, respectively, with variance $\sigma^2$ and zero mean. The received SINR for $x_1$ at $UE_2$ can be derived as follows:
\begin{equation}
\gamma_{x_1}^{t_1}={\rho{{|h_{S,2}|}^2}},
\end{equation}
where $\rho \triangleq \frac{P}{\sigma^2}$ denotes the transmit signal-to-noise ratio (SNR) and $\sigma^2$ denotes the additive white Gaussian noise variance for all received signals described in this study [20]. Following the principles of DL NOMA, during the second segment of phase 1 (($(1-\theta)T/3$)), $S$ transmits the superposition signal ($A_2$) to $UE_1$ and $UE_2$, as indicated below:
\begin{equation}
A_2=\sqrt{p_{N}P}x_2+\sqrt{p_{F}P}x_3, 
\end{equation}
where $x_2$ and $x_3$ denote the data symbol for $UE_1$ and $UE_2$, respectively [11--12]. $UE_1$ acquires $x_2$ using SIC. The imperfect SIC is assumed at $UE_1$. Thus, the residual interference due to imperfect SIC is quantified by $\beta$ ($0 \leq \beta \leq 1$). So, $\beta=1$ refers to imperfect SIC and $\beta=0$ refers as perfect SIC [22]. So, the received signals at $UE_1$ can be expressed by the following equation: 
\begin{equation}
y_{{S,1}_{NOMA}}^{t_1} = (\sqrt{p_{N}(1-\delta)P}x_2+\beta\sqrt{p_{F}(1-\delta)P}x_3)h_{S,1} + n_1.
\end{equation}
The received SINR corresponding to the symbols, $x_2$ ($\gamma_{x_2}^{t_1}$) and $x_3$ ($\gamma_{x_3->x_2}^{t_1}$) at $UE_1$ by SIC can be expressed as follows [11-13,23]:
\begin{equation}
\gamma_{x_2}^{t_1}=\frac{{(1-\delta)\rho{{|h_{S,1}|}^2}p_{N}}}{(1-\delta)+1}.
\end{equation}
\begin{equation}
\gamma_{x_3->x_2}^{t_1}=\frac{(1-\delta)\rho{{|h_{S,1}|}^2}p_F}{\beta^2(1-\delta)\rho{{|h_{S,1}|}^2}p_{N}+1}.
\end{equation}
 Furthermore, IHS-based SWIPT is used at $UE_1$ [11--13, 23]. According to Figure  2, $UE_1$ uses the $\delta$ proportion of the received power for EH. Moreover, $UE_1$ uses the fraction ($1-\delta$) of the received power for information decoding (ID). In addition, the directly received signal at $UE_2$ from $S$ is expressed as follows: 
 \begin{equation}
y_{{S,2}_{NOMA}}^{t_1}  = (\sqrt{p_{N}P}x_2+\sqrt{p_{F}P}x_3)h_{S,2} + n_2.
\end{equation}
 Furthermore, $x_3$ is directly decoded by $UE_2$ using the direct link from $S$. The received SINR corresponding to $x_3$ ($\gamma_{x_3}^{t_1}$) at $UE_2$ can be expressed as follows:
\begin{equation}
\gamma_{x_3}^{t_1}=\frac{\rho{{|h_{S,2}|}^2}p_F}{\rho{{|h_{S,2}|}^2}p_{N}+1}.
\end{equation}
\subsection{Phase-2($t_2$)}
During phase 2 (($(1-\theta)T/3$), $\hat x_3$ is transmitted by $P_{1}$ from $UE_1$ to $UE_2$ by utilizing the harvested energy (Figure  2). It can be assumed that $UE_1$ can perfectly decode the $x_3$ symbol during the second segment of phase 1 (($(1-\theta)T/3$)) (Figure  2) [11, 20, 23--24]. The power splitting (PS)-based SWIPT is implemented by $UE_1$ for EH to relay the decoded $\hat x_3$ from $UE_1$ to $UE_2$. The signal received by the DF relaying from $UE_1$ to $UE_2$ can be expressed as follows: 
\begin{equation}
y_{1,2}^{t_2} = \sqrt{P_{1}} \hat{x_3}h_{1,2} + n_2.
\end{equation}
Thus, the received SINR at $UE_2$ from $UE_1$ corresponding to the symbol $\hat x_3$ ($\gamma_{x_3}^{t_2}$) owing to the DF relay can be expressed as follows : 
\begin{equation}
\gamma_{x_3}^{t_2}={{{|h_{1,2}|}^2}P_{1}}.
\end{equation}
Thus, the DF relaying from $UE_1$ to $UE_2$ is performing by the transmitted power $P_1$ from $UE_1$. $P_1$ utilizing the harvested energy which can be expressed as follows by [11--13]: 
\begin{equation}
P_1= \frac{E_{1}}{(1-\theta)T/3} =\eta \rho |h_{S,1}|^2(\frac{3 \theta}{1-\theta}+\delta),
\end{equation}
where $E_{1}=\eta\rho|h_{S,1}|^2 \theta T + \eta \delta \rho |h_{S,1}|^2 (1-\theta) T/3 $ is the harvested energy at $UE_1$ by IHS protocol [11-13]. Moreover, $\eta$ represents the energy conversion efficiency that relies on the EH circuit and $0<\eta<1$ [11--13,20]. Moreover, $S$ directly transmitted a signal ($A_3=\sqrt{p_NP}x_4$) from $S$ to $UE_1$ by $p_NP$ during $t_2$. Because, $x_4$ is only transmitted for $UE_1$. So, the signal received at $UE_1$ during $t_2$ can be expressed as follows: 
\begin{equation}
y_{S,1}^{t_2} =(\sqrt{p_NP} x_4)h_{S,1} + n_1.
\end{equation}
Thus, the received SINR for $x_4$ at $UE_1$ during phase 2 owing to direct transmission can be represented as follows: 
\begin{equation}
\gamma_{x_4}^{t_2} = {p_N|h_{S,1}|^2 \rho}.
\end{equation}
\subsection{Channel Capacities of CNOMA-IHS}
By assuming normalized total time duration and total transmit power, the capacity of $x_1$ can be calculated as follows: 
\begin{equation}
C_{x_1}={\theta}\log_2(1+(\gamma_{x_1}^{t_1})).
\end{equation}
Moreover, the achievable capacities of $x_2$ and $x_3$ can be calculated as follows [2,9,11--13,20]:
\begin{equation}
C_{x_2}=\frac{1-\theta}{3}(\log_2(1+(\gamma_{x_2}^{t_1})).
\end{equation}
\begin{equation}
C_{x_3}=\frac{1-\theta}{3}(\log_2(1+min(\gamma_{x_3->x_2}^{t_1},\gamma_{x_3}^{t_1},\gamma_{x_3}^{t_2}))).
\end{equation}
In addition, the achievable capacities of $x_4$ can be calculated as follows [2,9,11--13]:
\begin{equation}
C_{x_4}=\frac{1-\theta}{3}\log_2(1+(\gamma_{x_4}^{t_2}).
\end{equation}
Thus, the achievable SC can be calculated using the following equations [2,9,11--13,20]:
\begin{equation}
C_{1} = \text{E}[C_{x_2}]+\text{E}[C_{x_4}]. 
\end{equation}
\begin{equation}
C_2 =\text{E}[C_{x_1}]+\text{E}[C_{x_3}]. 
\end{equation}
\begin{equation}
C_{sum} = C_{1}+C_{2},
\end{equation}
where E[.] represents the mean or expectation operator. Moreover, $C_{1}$, $C_{2}$, and $C_{sum}$ denote the respective channel capacities of $UE_1$, $UE_2$, and ESC for the proposed CNOMA-IHS scheme. 
\par
The analytical ergodic capacity (EC) of $UE_1$ can be expressed by Theorem 1. The values of the variables are considered identical. Various variables are assumed for information transfer by transmitting various symbols (i.e., $x_2$ and $x_4$) from $S$ in different phases for $UE_1$, utilizing different power allocations from $S$ in the case of the proposed scheme. The EC in the case of $UE_1$ for the proposed scheme can be analytically expressed using Theorem 1.\\
\textbf{Theorem 1.} The EC of $UE_1$ ($C_{1}^{erg}$) in the case of CNOMA-IHS is expressed by the following equation: 
\begin{equation}
C_{1}^{erg} = \frac{1-\theta}{3ln 2} \{ -Ei(\frac{-1}{g})e^{\frac{1}{g}} \}+\frac{1-\theta}{3ln 2} \{ -Ei(\frac{-1}{h})e^{\frac{1}{h}} \}, 
\end{equation}
where $g \triangleq  (1-\delta)\lambda_{S,1} \rho p_N$ and $h \triangleq \lambda_{S,1} \rho p_N$ due to direct transmission of $x_2$ by superimposed signal along with PS based EH and direct transmission of $x_4$ during $t_1$ and $t_2$, respectively. Moreover, Ei(.) denotes the exponential integral function. \\
\begin{proof}
Let $A \triangleq (1-\delta)\rho |h_{S,1}|^2 p_N$ and $B \triangleq \rho |h_{S,1}|^2 p_N$ due to direct transmission of $x_2$ by superimposed signal along with PS based EH and direct transmission of $x_4$ during $t_1$ and $t_2$, respectively; the cumulative distributed function (CDF) of $A$ and $B$ can be determined by the following equations[20,25--27]:
\begin{equation}
F_a(A) = 1-e^{\frac{-a}{(1-\delta)\lambda_{S,1} p_N \rho}}, and
\end{equation}
\begin{equation}
F_b(B)=1-e^{\frac{-b}{\lambda_{S,1} p_N \rho}}. 
\end{equation}
The EC of $UE_1$ can be derived by determining $\int_{0}^{\infty} (1+a) f_A(a)da =  \frac{1}{ln2}\int_{0}^{\infty} \frac{1-F_A(a)}{1+a}da$ and $\int_{0}^{\infty} (1+b) f_B(b)db = \frac{1}{ln2}\int_{0}^{\infty} \frac{1-F_B(b)}{1+b}db$. Thus, the EC of $UE_1$ can be derived as follows :
\begin{multline*}    
C_{1}^{erg} = \frac{1-\theta}{3ln2}\int_{0}^{\infty} \frac{1}{1+a} e^{\frac{-a}{(1-\delta)\lambda_{S,1} \rho p_N}} da+ \frac{1-\theta}{3ln2}\int_{0}^{\infty} \frac{1}{1+b} e^{\frac{-b}{\lambda_{S,1} \rho p_N}}. \\
\end{multline*}
\begin{multline*}    
C_{1}^{erg} = \frac{1-\theta}{3ln2}\{ -Ei(\frac{-1}{(1-\delta)\lambda_{S,1} \rho p_N})e^{\frac{1}{(1-\delta)\lambda_{S,1} \rho p_N}} \}+\frac{1-\theta}{3ln2}\{ -Ei(\frac{-1}{\lambda_{S,1} \rho p_N})e^{\frac{1}{\lambda_{S,1} \rho p_N}} \}, 
\end{multline*}    
\begin{equation} 
=\frac{1-\theta}{3ln 2} \{ -Ei(\frac{-1}{g})e^{\frac{1}{g}} \}+\frac{1-\theta}{3ln 2} \{-Ei(\frac{-1}{h})e^{\frac{1}{h}} \}, 
\end{equation}
where $g \triangleq (1-\delta)\lambda_{S,1} \rho p_N$ and $h \triangleq \lambda_{S,1} \rho p_N$ due to direct transmission of $x_2$ by superimposed signal along with PS based EH and direct transmission of $x_4$ during $t_1$ and $t_2$, respectively. Moreover, Ei(.) represents the exponential integral function.
\end{proof}
The EC of $UE_2$ for the proposed CNOMA-IHS scheme can be analytically derived using Theorem 2 \\
\textbf{Theorem 2.} The EC of $UE_2$ ($C_{2}^{erg}$) for the CNOMA-IHS scheme is expressed as follows: 
\begin{equation} \label{eq27}
\begin{split}
C_{2}^{erg} & = \frac{1-\theta}{3ln2} \{ -Ei(-(l+q))e^{(l+q)} \\+ & Ei(-(r+s))e^{(r+s)} \}+\frac{\theta}{ln2} \{ -Ei( \frac{-1}{u})e^{ \frac{1}{u}} \}, 
\end{split}
\end{equation}
where, $l = \frac{1}{\lambda_{S,1} \rho (1-\delta)}$, $q = \frac{1}{\lambda_{1,2} \rho}$, $r = \frac{1}{\lambda_{S,1} \rho p_N(1-\delta)}$, $s = \frac{1}{\lambda_{1,2} \rho p_N }$, $u = {\lambda_{S,2} \rho}$, and Ei(.) represents the exponential integral function.\\ 
\begin{proof} Let $L\triangleq \frac{\rho{{|h_{S,1}|}^2}p_F(1-\delta)}{\rho{{|h_{S,1}|}^2}p_{N}(1-\delta)+1}$ by considering the imperfect SIC, $Q \triangleq \frac{\rho{{|h_{S,2}|}^2}p_F}{\rho{{|h_{S,2}|}^2}p_{N}+1}$, $M \triangleq min(\gamma_{x_3 \rightarrow x_2}^{t_1},\gamma_{x_3}^{t_1},\gamma_{x_3}^{t_2})$, and $V \triangleq {\rho{{|h_{S,2}|}^2}}$. Thus, the CDF of $L$, $Q$, $M$, and $V$ can be written as follows [20,25--27]: 
\begin{equation}
F_l(L) = 1-\frac{\rho \lambda_{S,1} p_F(1-\delta)}{(1-\delta)(\rho \lambda_{S,1} p_F+\rho \lambda_{S,1} p_N)}e^{\frac{-l}{\rho \lambda_{S,1} p_F (1-\delta)}}. 
\end{equation}
\begin{equation}
F_q(Q) = 1-\frac{\rho \lambda_{S,2} p_F}{\rho \lambda_{S,2} p_F+\rho \lambda_{S,2} p_N}e^{\frac{-q}{\rho \lambda_{S,2} p_F}}. 
\end{equation}
\begin{equation}
F_m(M) = (1-e^{\frac{-m}{(r+s)}})-(1-e^{\frac{-m}{(r+s)}}). 
\end{equation}
\begin{equation}
F_v(V) = 1-e^{\frac{-v}{\lambda_{S,2} \rho P}}. 
\end{equation}
Using $\int_{0}^{\infty} (1+m) f_M(m)dm = \frac{1}{ln2}\int_{0}^{\infty} \frac{1-F_m(M)}{1+m}dm$ and $\int_{0}^{\infty} (1+v) f_V(v)dv = \frac{1}{ln2}\int_{0}^{\infty} \frac{1-F_v(V)}{1+v}dv$, the EC of $UE_2$ can be written as (29). After the mathematical manipulation, the EC of the $UE_2$ can be achieved as below [27]:
\begin{equation} \label{eq27}
\begin{split}
C_{2}^{erg} & = \frac{1-\theta}{3ln2} \{ -Ei(-(l+q))e^{(l+q)} \\+ & Ei(-(r+s))e^{(r+s)} \}+\frac{\theta}{ln2} \{ -Ei( \frac{-1}{u})e^{ \frac{1}{u}} \}. 
\end{split}
\end{equation}
where, $l = \frac{1}{\lambda_{S,1} \rho(1-\delta)}$, $q = \frac{1}{\lambda_{1,2} \rho}$, $r = \frac{1}{\lambda_{S,1} \rho p_N (1-\delta)}$, $s = \frac{1}{\lambda_{1,2} \rho p_N}$, $u = {\lambda_{S,2} \rho}$, and Ei(.) represents the exponential integral function.
\end{proof}
By adding (25) and (29), the analytical expression of ESC of the proposed CNOMA-IHS can be derived as follows:
\begin{equation}
C_{sum}^{erg} = C_{1}^{erg}+C_{2}^{erg}.
\end{equation}

\subsection{Energy Efficiency of CNOMA-IHS}
The evaluation of EE and the optimization technique of EE for the proposed CNOMA-IHS scheme is describe in this subsection. $UE_1$ uses the energy harvested by the proposed CNOMA-IHS scheme to conduct a relay operation. The relay of $\hat{x_3}$ from $UE_1$ to $UE_2$ is conducted during phase 2 of the proposed IHS protocol using $P_{1}$. Thus, EE is the ratio of the ESC ($C_{sum}$) to the total transmit power for direct transmission ($2P$ and $p_N$) and transmit power of $UE_1$ for DF-based relay operation ($P_{1}$) [28]. Therefore, the EE corresponding to the proposed CNOMA-IHS scheme can be derived using the following equation:
\begin{equation}
EE=\frac{C_{sum}}{2P+P_{1}+p_N}. 
\end{equation}\\
As $\theta$ is the dominating factor than $\delta$ in case of HS SWIPT protocol [11--13]. Hence $\theta$ based EE optimization technique is considered for the proposed CNOMA-IHS scheme [18]. Thus, the optimal $\theta$ ($\theta^*$) can be derived to achieve EE maximization by below equation: 
\begin{equation}
\theta^{*}= 1-\frac{E_1}{\eta |h_{S,1}|^2 (2P+P_1+p_N)}. 
\end{equation}

\subsection{OMA-IHS}
As benchmark, OMA-IHS scheme is considered and compared with the proposed CNOMA-IHS scheme for fair comparisons. In the case of OMA, time division multiple access has been considered in this study. In this scenario, $S$ directly delivers different information signals to $UE_1$ and $UE_2$ separately using various independent time slots. Furthermore, one additional time slot is required to perform TS-based EH at $UE_1$. Moreover, an additional time slot is required to perform the DF relay of $\hat x_3$ from $UE_1$ to $UE_2$. Various independent time slots are allocated for $UE_1$ and $UE_2$ related to various symbols for information transfer (e.g., $x_1$, $x_2$, $x_3$, and $x_4$), EH (e.g., EH at $UE_1$) and for DF relaying of $\hat x_3$ are denoted by $t_1$, $t_2$, $t_3$, $t_4$, $t_5$, and $t_6$, respectively, because the IHS protocol is a combination of TS and PS-based SWIPT. Total six time slots are used in case of OMA-IHS scheme. Among them, four time slots are used for information transfer from $S$ to the users and one time slot is using for TS-based EH and another for DF relaying. At first, $x_1$ is transmitted to $UE_1$ during $t_1$ for the TS-based EH. Then, $x_1$ is transmitted to $UE_2$ during $t_2$ for information transfer. In addition, $x_2$ is transmitted from $S$ to $UE_1$ by $t_3$ for information transfer. In the case of PS-based SWIPT, $UE_1$ uses a fraction ($\delta$) of the received power for EH. Furthermore, the rest of the fraction ($1-\delta$) of the received power is used for ID during time slot $t_3$ [17, 20]. Subsequently, $x_3$ is transmitted from $S$ to $UE_2$ by $t_4$ for information transfer. Moreover, $\hat x_3$ relays from $UE_1$ to $UE_2$ during $t_5$ using the power ($P_1$) which utilizing the harvested energy. Afterwards, $x_4$ is directly transmitted to $UE_1$ from $S$ during $t_6$ for information transfer as well. All these transmissions are performed by the total transmit power of $P$ from $S$. In addition, $t_1=t_2=t_3=t_4=t_5=t_6=\frac{1}{6}$ are considered in this study for the OMA-IHS scheme to complete all the EH, information transfer, and relaying. Thus, the capacity of $x_1$ at $UE_2$ in the case of OMA-IHS can be expressed as follows [2,20,25--26]:
\begin{equation}
C_{x_1}^{OMA}=\frac{1}{6}(\log_2(1+(\gamma_{x_1}^{OMA}))).
\end{equation}
In addition, the achievable capacities of $x_2$ and $x_3$ corresponding to the OMA-IHS scheme can be calculated as follows [2,20,24,26]:
\begin{equation}
C_{x_2}^{OMA}=\frac{1}{6}(\log_2(1+(\gamma_{x_2}^{OMA}))),
\end{equation}
\begin{equation}
C_{x_3}^{OMA}=\frac{1}{6}(\log_2(1+min(\gamma_{x_3}^{OMA},\gamma_{\hat x_3}^{OMA}))), and
\end{equation}
\begin{equation}
C_{x_4}^{OMA}=\frac{1}{6}(\log_2(1+(\gamma_{x_4}^{OMA}))),
\end{equation}
where
$\gamma_{x_1}^{OMA}={\rho{{|h_{S,2}|^2P}}}$. 
$\gamma_{x_2}^{OMA}={(1-\delta)\rho{{|h_{S,1}|}^2}P}$, 
$\gamma_{x_3}^{OMA}={(1-\delta)\rho{{|h_{S,1}|}^2}P}$,
$\gamma_{x_4}^{OMA}=\rho{{{|h_{S,1}|}^2}P}$, 
$\gamma_{\hat x_3}^{OMA}={\rho{{|h_{1,2}|}^2}P_{1}^{OMA}}$, $\gamma_{x_4}^{OMA}={\rho{{|h_{1,2}|}^2}P}$, and 
$P_{1}^{OMA}=\eta \rho |h_{S,1}|^2(\frac{6 \theta}{1-\theta}+\delta)$.
Thus, the ESC of the OMA-IHS scheme can be expressed as follows [2,20,25--26]:
\begin{equation}
C_{1}^{OMA} = {E}[C_{x_2}^{OMA}]+{E}[C_{x_4}^{OMA}]. 
\end{equation}
\begin{equation}
C_{2}^{OMA} ={E}[C_{x_1}^{OMA}]+{E}[C_{x_3}^{OMA}]. 
\end{equation}
\begin{equation}
C_{sum}^{OMA} =C_{1}^{OMA}+C_{2}^{OMA}. 
\end{equation}
Here, E[.] denotes the expectation operator. $C_{1}^{OMA}$, $C_{2}^{OMA}$, and $C_{sum}^{OMA}$ denote the respective channel capacities of $UE_1$, $UE_2$, and ESC in case of the OMA-IHS scheme. Moreover, the associated EE for the OMA-IHS can be derived using the following equation [13,15,28]:
\begin{equation}
EE_{OMA}=\frac{C_{sum}^{OMA}}{5P+P_{1}^{OMA}}. 
\end{equation}
Therefore, the aforementioned equation demonstrates that EE is related to $C_{sum}^{OMA}$, $P$, and $P_{1}^{OMA}$.
\section{Result Analysis}

The results of the ESC of the proposed CNOMA-IHS scheme and the compared existing schemes (e.g., CNOMA-HS[11], WP-CNOMA[14], CNOMA-PS[20], CNOMA-TS[20], and OMA-IHS) are evaluated in this section. All the simulation result evaluations were performed using MATLAB. Moreover, the impacts of $\theta$, $\delta$, and $\eta$ are examined for the proposed CNOMA-IHS scheme and other compared schemes. In addition, the impact of the transmit SNR ($\rho$) and the distance between $S$ and $UE_1$ ($d_{S,1}$) on ESC in case of the proposed scheme and other compared schemes have been evaluated. A comparative analysis in terms of EE corresponding to the proposed CNOMA-IHS scheme and other compared schemes has been presented. The EE of the proposed CNOMA-IHS scheme with the EE optimization technique and without EE optimization technique are compared with other existing schemes is also presented in this section. The impact of $\theta$ on the EE in case of the proposed CNOMA-IHS scheme and other compared schemes have been also evaluated in this section. It should be noted that similar simulation parameters have been applied to the proposed and other compared schemes.
\subsection{Ergodic Sum Capacity (ESC)}

\begin{figure}[h!]
\centering
\includegraphics[width=0.6\textwidth]{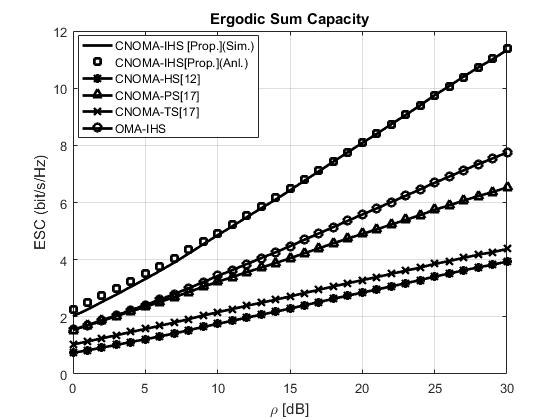}
\caption{Comparisons of ESC versus $\rho$.}
\label{image-myimage}
\end{figure}

\begin{figure}[h!]
\centering
\includegraphics[width=0.6\textwidth]{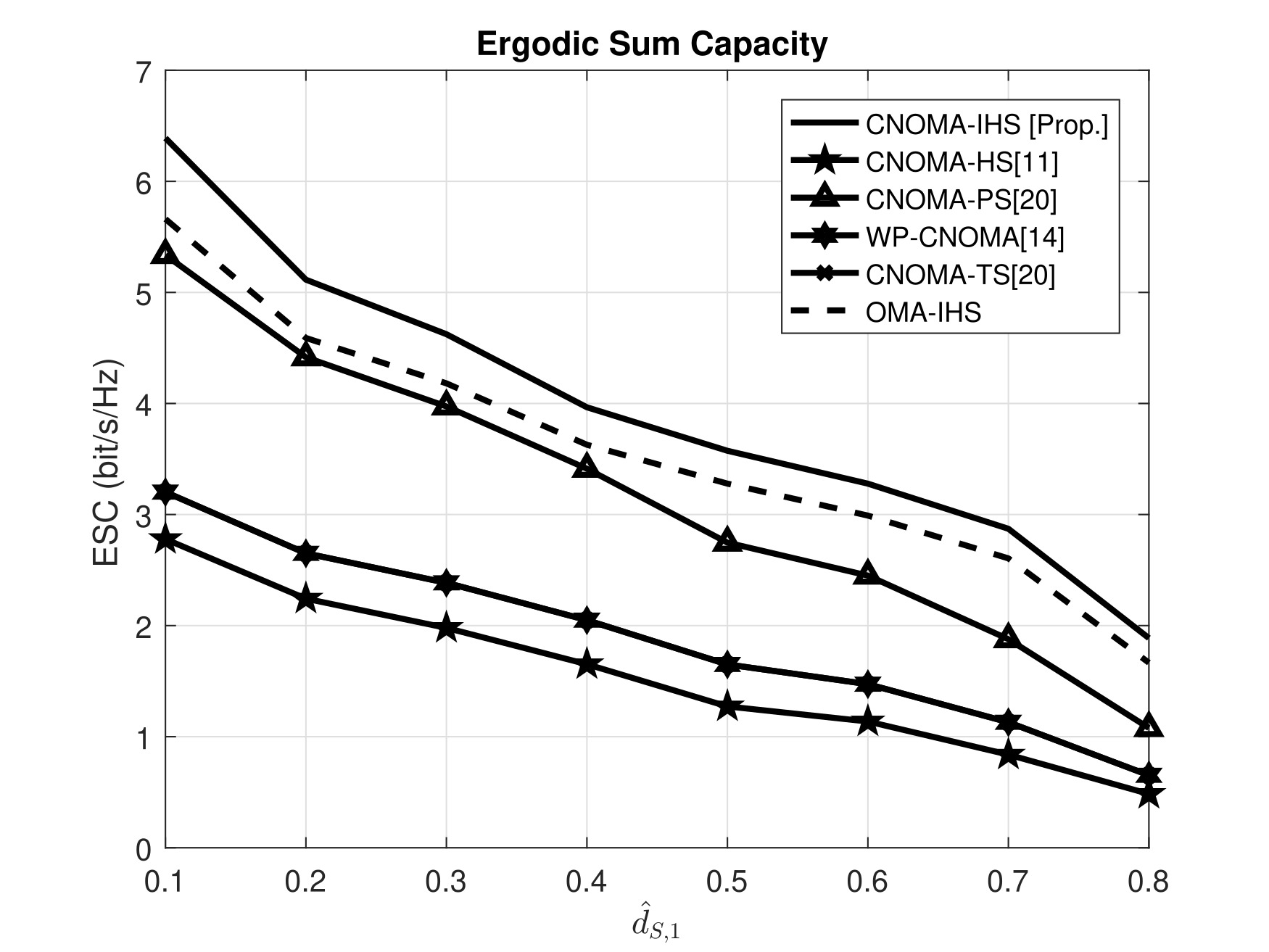}
\caption{Comparisons of ESC versus $d_{S,1}$.}
\label{image-myimage}
\end{figure}

\begin{figure}[h!]
\centering
\includegraphics[width=0.6\textwidth]{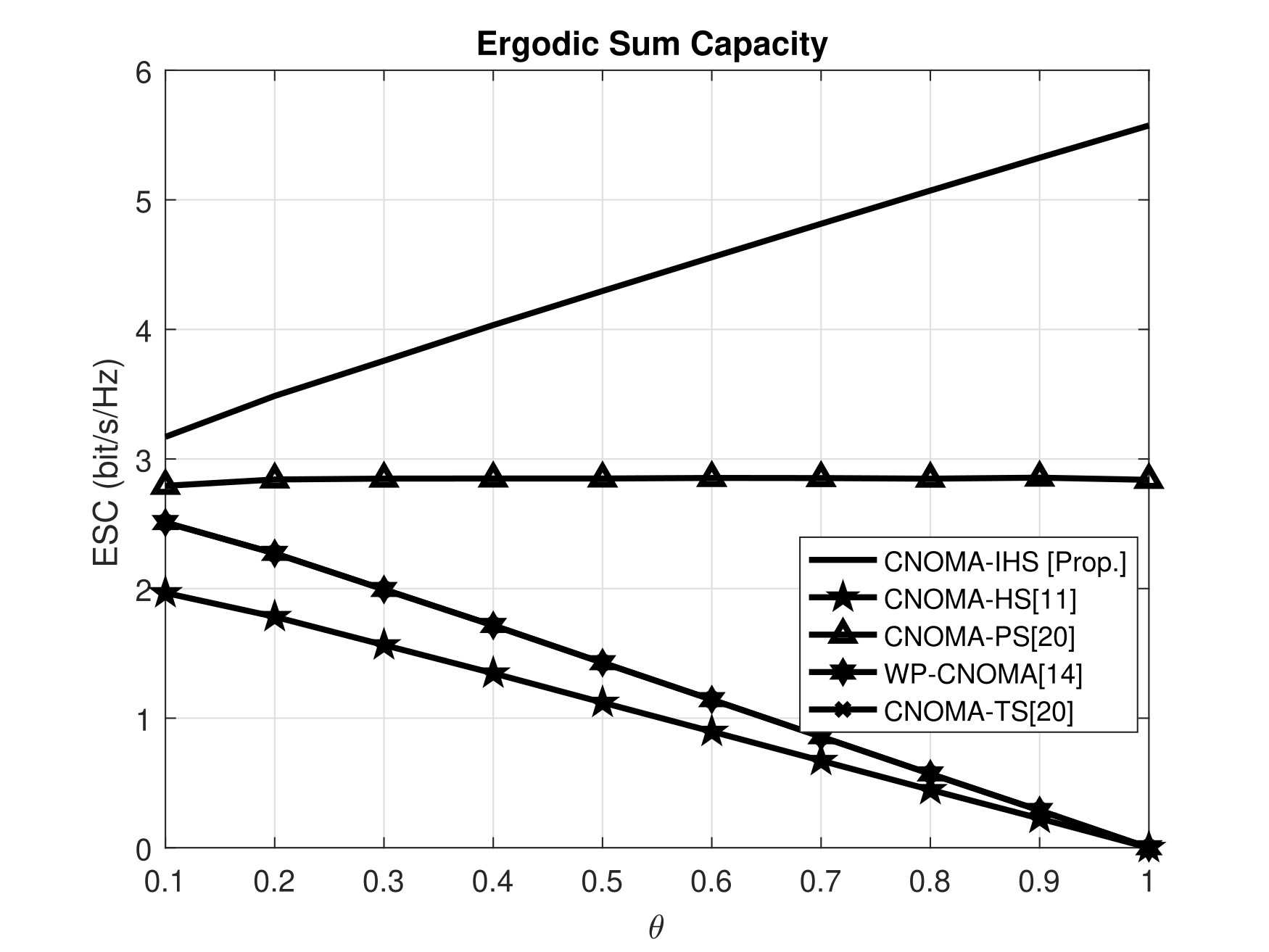}
\caption{ Impact of $\theta$ on the ESC.}
\label{image-myimage}
\end{figure}

\begin{figure}[h!]
\centering
\includegraphics[width=0.6\textwidth]{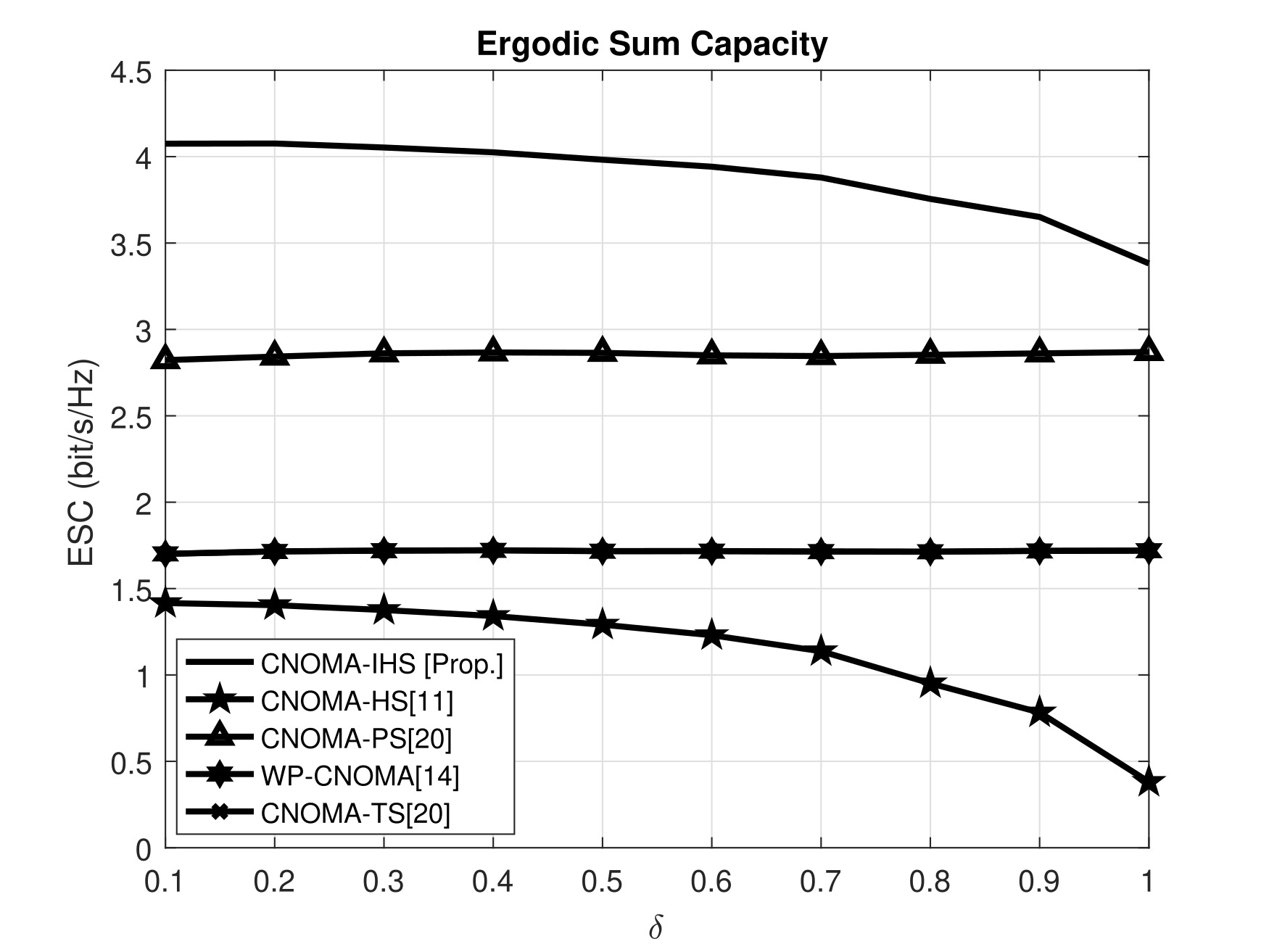}
\caption{Impact of $\delta$ on the ESC.}
\label{image-myimage}
\end{figure}

\begin{figure}[h!]
\centering
\includegraphics[width=0.6\textwidth]{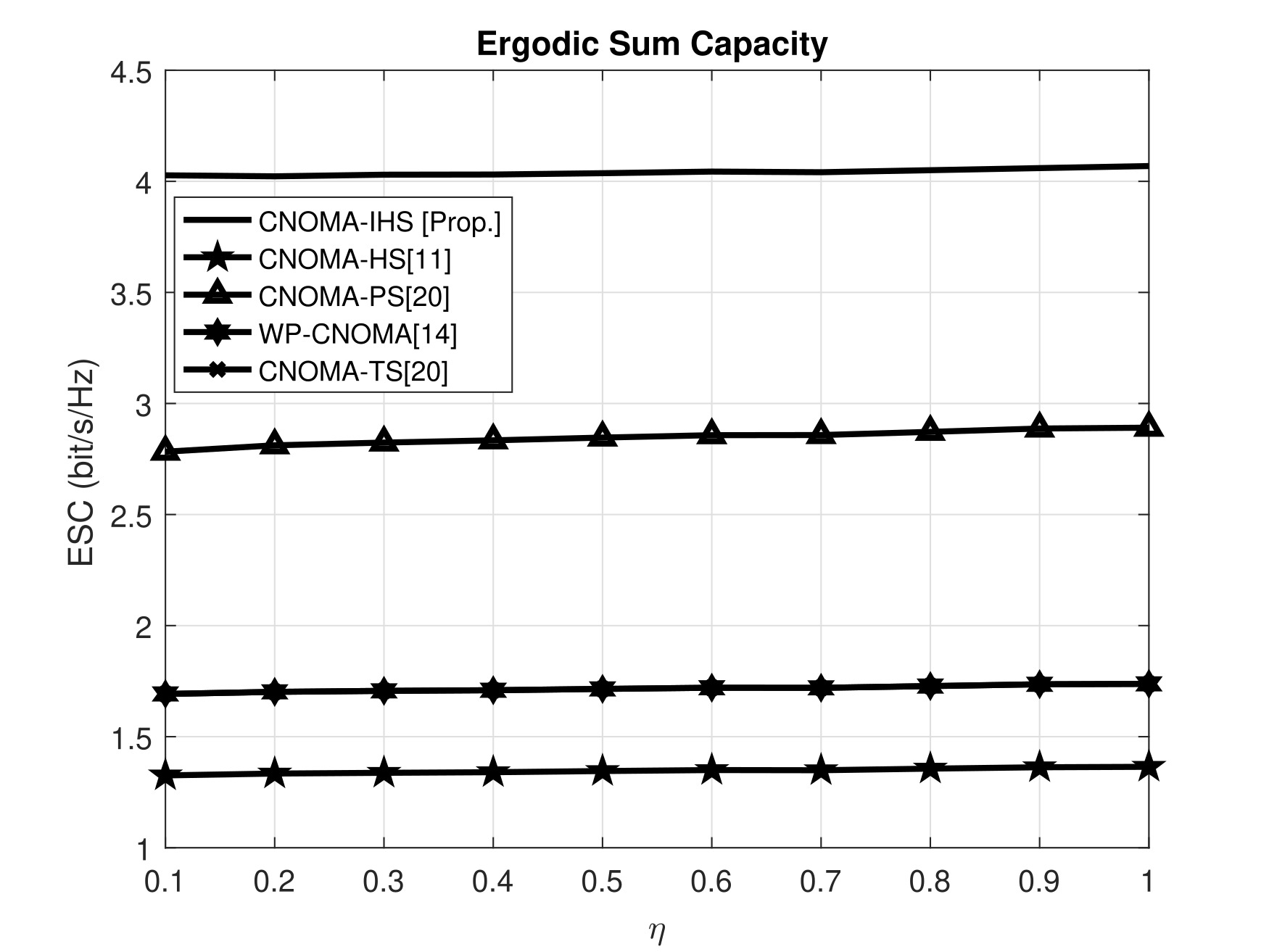}
\caption{ Impact of $\eta$ on the ESC.}
\label{image-myimage}
\end{figure}

\begin{figure}[h!]
\centering
\includegraphics[width=0.6\textwidth]{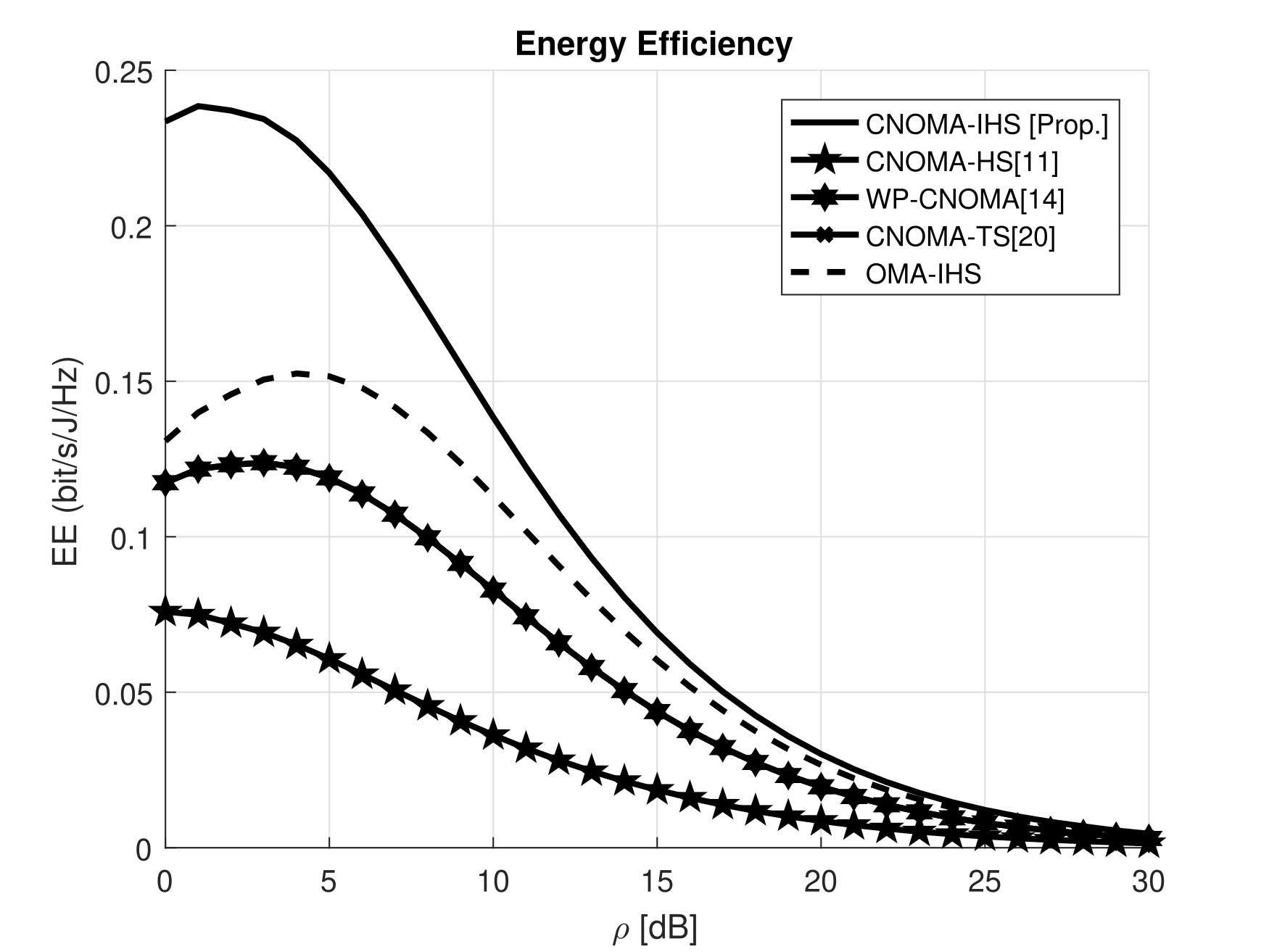}
\caption{Comparisons of EE with respect to $\rho$.}
\label{image-myimage}
\end{figure}

\begin{figure}[h!]
\centering
\includegraphics[width=0.6\textwidth]{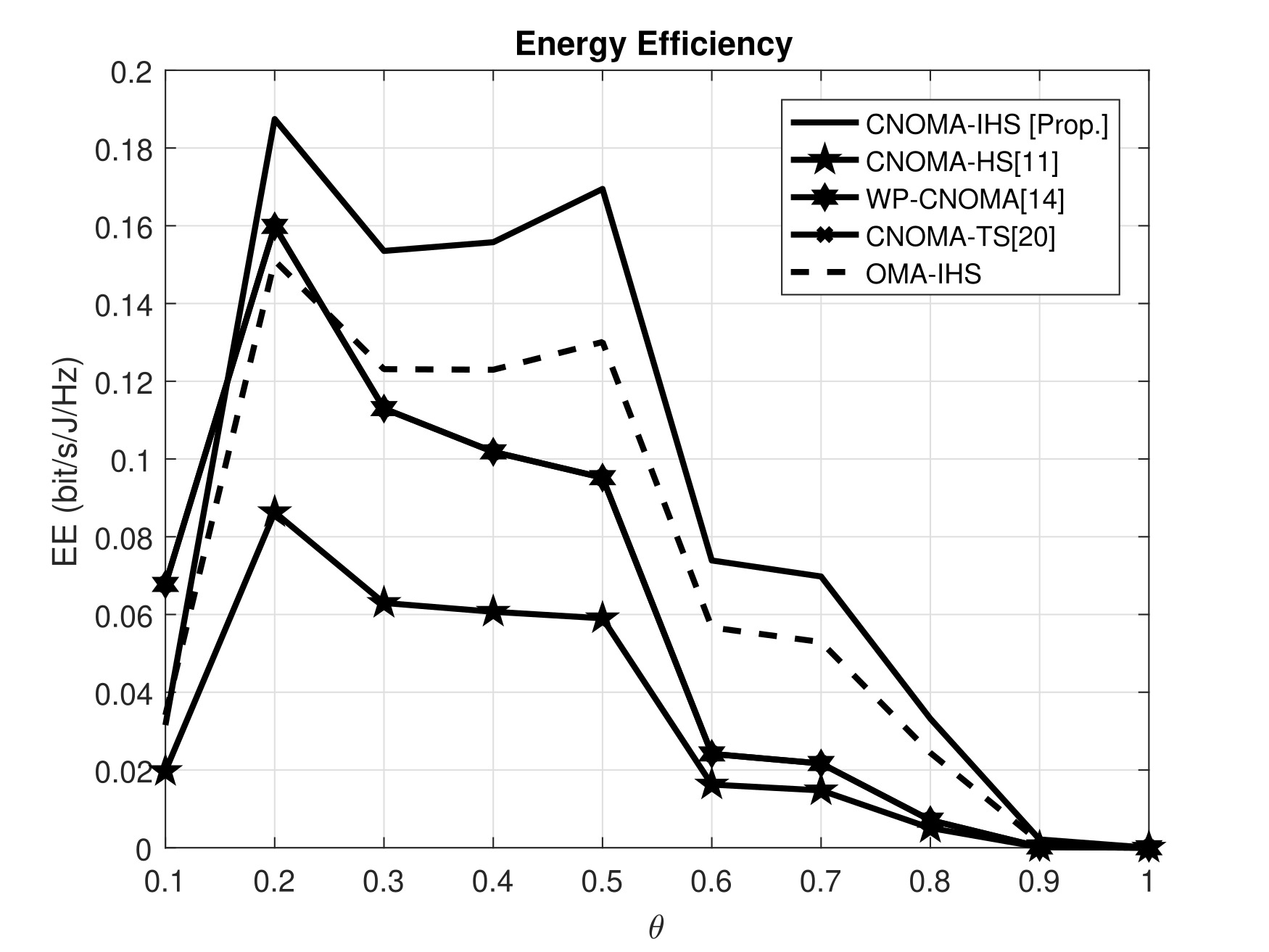}
\caption{Comparisons of EE with respect to $\theta$.}
\label{image-myimage}
\end{figure}

\par
Figure 3 illustrates that the proposed scheme exhibits a higher capacity than the conventional schemes for different transmit SNRs ($\rho$). The parameters, $T=1$, $d_{S,1}=0.6$, $d_{S,2}=1$, $\eta=1$, $\delta=\theta=0.4$, $R_{th,1}=R_{th,2}=0.3$, $\beta=1$, $v=2$, and $d_{1,2}=d_{S,2}-d_{S,1}$, are considered for the MATLAB simulation. Moreover, for each value of $\rho$, the ESC of the proposed CNOMA-IHS scheme is significantly higher than those of the other schemes. This is because the additional symbols ($x_1$ and $x_4$) are transmitted to $UE_2$ and $UE_1$, respectively, using the CNOMA-IHS scheme without consuming any additional resource. Hence, the user channel capacities and ESC improvement is achieved for the proposed scheme. Thus, the proposed scheme exhibits higher ESC than the other compared schemes. Moreover, the analytical results validate the simulation results of ESC for the proposed CNOMA-IHS scheme which depicted in Figure 3.      
\par
Figure 4 illustrates that owing to an increase in $d_{S,1}$, the ESC corresponding to the proposed scheme is higher than those corresponding to other compared schemes. The parameters, $T=1$, $d_{S,2}=1$, $\eta=1$, $\delta=\theta=0.4$, $\rho=15dB$, $R_{th,1}=R_{th,2}=0.3$, $\beta=1$, $v=2$, and $d_{1,2}=d_{S,2}-d_{S,1}$, are considered for the simulation. The position of the relaying node ($UE_1$) varies depending on the variations of $d_{S,1}$. But $x_1$ and $x_4$ are transmitted towards $UE_1$ and $UE_2$ during phase-1 and phase-2, respectively. Thus, the achieved ESC is significantly higher than the other compared schemes. Furthermore, in the case of the proposed scheme, the amount of energy harvested is higher than those of the other schemes, as CNOMA-IHS comprises PS- and TS-based SWIPT protocols. Thus, the harvested energy is sufficient for the relay operation by $UE_1$ in the case of higher values of $d_{S,1}$ corresponding to the proposed CNOMA-IHS scheme.
\par
The comparative analysis depicted in Figure 5 indicates that increasing $\theta$ influences the ESC of the proposed CNOMA-IHS scheme. The parameters, $T=1$, $d_{S,1}=0.6$, $d_{S,2}=1$, $\rho=15 dB$, $\delta=0.4$, $\eta=1$, $R_{th,1}=R_{th,2}=0.3$, $\beta=1$, $v=2$, and $d_{1,2}=d_{S,2}-d_{S,1}$, are considered for the MATLAB simulation. For increasing values of $\theta$, ESC decreases significantly in the case of conventional TS-SWIPT protocol based schemes (CNOMA-HS [11], WP-CNOMA [14], and CNOMA-TS [20]) but ESC increases significantly in the case of the proposed CNOMA-IHS scheme. Because of the higher value of $\theta$ can enhance the harvested energy at $UE_1$ for relaying. Hence, only $\theta T$ segment of phase-1 is directly involved to perform TS-based EH at $UE_1$ in the case of conventional TS-SWIPT protocol based schemes (CNOMA-HS[12] (HS protocol is the combination of TS and PS based SWIPT), WP-CNOMA[22], and CNOMA-TS[20]). Hence, the remaining time segments are not sufficient which are used for NOMA-based direct transmission, and DF relaying [11--13,20,23]. Therefore, higher values of $\theta$ degrade the user capacities of the conventional TS-based SWIPT schemes. Hence, the ESC of the CNOMA-HS [12], WP-CNOMA [14], and CNOMA-TS [20] schemes, is degraded owing to the increasing values of $\theta$. In contrast, $x_1$ is transmitted to $UE_2$ during the first segment of phase-1 ($\theta T$) along with the TS-based EH at $UE_1$, which enhances the capacity of $UE_2$ as well as ESC of the proposed CNOMA-IHS scheme. Though other time segments of the protocol for the proposed CNOMA-IHS scheme are not sufficient to perform direct transmission and DF relaying. Thus, the proposed CNOMA-IHS scheme provides higher ESC than other compared TS-SWIPT based CNOMA schemes owing to increasing values of $\theta$.
\par
The comparative analysis illustrated in Figure 6 indicates that $\delta$ influences the ESC of the proposed CNOMA-IHS scheme. The parameters,$T=1$, $d_{S,1}=0.6$, $d_{S,2}=1$, $\rho=15 dB$, $\theta=0.4$, $\eta=1$, $R_{th,1}=R_{th,2}=0.3$, $\beta=1$, $v=2$, and $d_{1,2}=d_{S,2}-d_{S,1}$, are considered for the MATLAB simulation. The ESC is decreasing in the case of the proposed and PS-based schemes (CNOMA-IHS, CNOMA-HS[12], and CNOMA-PS[20]) for increasing values of $\delta$. Due to the higher value of $\delta$, the harvested energy at $UE_1$ is increase but the ID cannot perform successfully because ($1-\delta$) amount of energy is used for ID at $UE_1$ which is degraded due to the higher value of $\delta$. Hence, the relaying cannot perform effectively by $UE_1$ since the signal decoding cannot perform successfully by less amount of energy due to the higher value of $\delta$. Furthermore, in the case of the proposed CNOMA-IHS scheme, additional symbols ($x_1$ and $x_4$) are transmitted to $UE_2$ and $UE_1$ without consuming any additional resources or interference. Hence, the higher individual user channel capacities and higher ESC are also achieved for the proposed scheme due to different values of $\delta$.
\par
The comparative analysis depicted in Figure  7 indicates the impact of $\eta$ on the ESC of the proposed scheme and the other conventional schemes. \textcolor{blue}{The parameters, $T=1$, $d_{S,1}=0.6$, $d_{S,2}=1$, $\rho=15 dB$, $\delta=\theta=0.4$, $R_{th,1}=R_{th,2}=0.3$, $\beta=1$, $v=2$, and $d_{1,2}=d_{S,2}-d_{S,1}$, are considered for the MATLAB simulation.} Figure 7 illustrates that $\eta$ does not have much influence on the ESC in the case of the proposed scheme and other compared schemes. However, CNOMA-IHS transmits the additional symbols ($x_1$ and $x_4$) to the users utilizing different phases without consuming any additional resources or any interference issue. Hence, the user channel capacities, as well as ESC, corresponding to the proposed CNOMA-IHS scheme are significantly higher compared with the other schemes for different values of $\eta$.
\par

\subsection{Energy Efficiency (EE)}

The EE comparisons between the proposed CNOMA-IHS scheme and the other compared schemes are depicted in Figure 8 with and without the EE optimization technique. The parameters, $T=1$, $d_{S,1}=0.5$, $d_{S,2}=1$, $\eta=1$, $\delta=\theta=0.4$, $\rho=15dB$, $R_{th,1}=R_{th,2}=0.3$, $\beta=1$, $v=2$, and $d_{1,2}=d_{S,2}-d_{S,1}$, are considered for the MATLAB simulation. The CNOMA-IHS exhibits higher EE than other TS-based compared schemes (e.g., CNOMA-HS[11], WP-CNOMA[14], CNOMA-TS[20], and OMA-IHS). This phenomenon occurs because additional symbols ($x_1$ and $x_4$) are transmitted to the users by the proposed CNOMA-IHS scheme to enhance the ESC of the proposed compared scheme. This leads to the use of additional power by the proposed scheme but significantly higher ESC is also achieved by the proposed CNOMA-IHS scheme. Thus, the proposed scheme outperforms the other schemes in case of EE without EE optimization. In addition, due to the optimized value of $\theta$ ($\theta^*$), significant EE gain can be achieved in the case of the proposed CNOMA-IHS scheme with EE optimization compared to the proposed scheme without EE optimization. \par
The comparative analysis depicted in Figure 9 indicates that increasing $\theta$ influences the EE of the proposed CNOMA-IHS scheme. The parameters, $T=1$, $d_{S,1}=0.6$, $d_{S,2}=1$, $\rho=15 dB$, $\delta=0.4$, $\eta=1$, $R_{th,1}=R_{th,2}=0.3$, $\beta=1$, $v=2$, and $d_{1,2}=d_{S,2}-d_{S,1}$, are considered for the MATLAB simulation. For increasing values of $\theta$, EE decreases significantly in the case of the proposed CNOMA-IHS scheme and conventional TS-SWIPT protocol-based schemes (CNOMA-HS [11], WP-CNOMA [14], and CNOMA-TS [20]) schemes. Because the higher value of $\theta$ can enhance the harvested energy at $UE_1$ for relaying. Hence, only $\theta T$ segment of phase-1 is directly involved to perform TS-based EH at $UE_1$ in the case of the proposed CNOMA-IHS scheme and other conventional TS-SWIPT protocol based schemes as well (CNOMA-HS[12] (HS protocol is the combination of TS and PS based SWIPT), WP-CNOMA[22], and CNOMA-TS[20]). For increasing values of $\theta$, EE decreases significantly in the case of all SWIPT protocol-based schemes (CNOMA-IHS [Prop.], CNOMA-HS [11], WP-CNOMA [14], and CNOMA-TS [20]). Because a higher value of $\theta$ can enhance the harvested energy at $UE_1$ for relaying. So, higher transmitted power is utilized a higher amount of harvested energy for the same amount of ESC. Moreover, in the case of the proposed CNOMA-IHS scheme the transmitted power is higher cause $x_4$ is transmitted during phase-2 with $p_N$ compared to conventional CNOMA-HS scheme. But the proposed CNOMA-IHS scheme provides significantly higher EE than other conventional schemes because higher ESC is achieved in case of the CNOMA-IHS scheme compared to other existing schemes.

\section{Conclusion}
In this study, a CNOMA-IHS scheme was proposed, and its performance was evaluated in terms of ESC and EE. The CCU was used as a DF relay for the CEU in the proposed scheme. The ESC of the proposed CNOMA-IHS scheme was evaluated and compared with CNOMA and existing SWIPT protocol-based schemes (e.g., CNOMA-HS[11], CNOMA-PS[20], WP-CNOMA[14], and CNOMA-TS[20] and the OMA-IHS scheme). The advantages of the proposed CNOMA-IHS scheme over other compared schemes were demonstrated based on the evaluation of the analytical and simulation results. To evaluate the system performance in terms of ESC, the impact of different parameters (e.g., $\delta$, $\theta$, $\eta$, $\rho$, and $d_{S,1}$) on the ESC was investigated. The proposed CNOMA-IHS scheme was superior to the conventional SWIPT based schemes in terms of ESC, as evidenced by the simulation and analytical results. Furthermore, the EE of the proposed CNOMA-IHS scheme was higher than that of the conventional TS-SWIPT based schemes (e.g., CNOMA-HS[11], WP-CNOMA [14], CNOMA-TS[20], and OMA-IHS). Moreover, the impact of $\theta$ on EE in the case of the proposed scheme and other compared schemes are also illustrated by simulation results. In addition, due to the considered EE optimization technique with the CNOMA-IHS scheme provide significantly higher EE than the proposed CNOMA-IHS without EE optimization technique and existing TS-SWIPT based schemes. This study can be extended by integrating the multi-antenna-based BS and transmit antenna selection technique with the proposed CNOMA-IHS scheme.


%

\section*{Acknowledgment}

This work was supported by the National Research Foundation of Korea (NRF) funded by the Korean government (MSIT) under grant No. 2019R1A2C1089542.

\ifCLASSOPTIONcaptionsoff
  \newpage
\fi


\begin{thebibliography}{00}

\bibitem{b1}L. Dai, B. Wang, Z. Ding, Z. Wang, S. Chen and L. Hanzo, "A Survey of Non-Orthogonal Multiple Access for 5G," in IEEE Communications Surveys \& Tutorials, vol. 20, no. 3, pp. 2294-2323, thirdquarter 2018.

\bibitem{b2} M. B. Shahab, M. F. Kader and S. Y. Shin, "A Virtual User Pairing Scheme to Optimally Utilize the Spectrum of Unpaired Users in Non-orthogonal Multiple Access," in IEEE Signal Processing Letters, vol. 23, no. 12, pp. 1766-1770, Dec. 2016.

\bibitem{b3} Wan, M. Wen, X. Cheng, S. Mumtaz and M. Guizani, "A Promising Non-Orthogonal Multiple Access Based Networking Architecture: Motivation, Conception, and Evolution," in IEEE Wireless Communications, vol. 26, no. 5, pp. 152-159, October 2019.

\bibitem{b4} A. A. Amin and S. Y. Shin, "Channel Capacity Analysis of Non-Orthogonal Multiple Access With OAM-MIMO System," in IEEE Wireless Communications Letters, vol. 9, no. 9, pp. 1481-1485, Sept. 2020.

\bibitem{b5} G. Song and X. Wang, "Comparison of Interference Cancellation Schemes for Non-Orthogonal Multiple Access System," 2016 IEEE 83rd Vehicular Technology Conference (VTC Spring), Nanjing, 2016, pp. 1-5.

\bibitem{b6} S. M. R. Islam, N. Avazov, O. A. Dobre and K. Kwak, "Power-Domain Non-Orthogonal Multiple Access (NOMA) in 5G Systems: Potentials and Challenges," in IEEE Communications Surveys \& Tutorials, vol. 19, no. 2, pp. 721-742, Secondquarter 2017. 


\bibitem{b7} Z. Ding, X. Lei, G. K. Karagiannidis, R. Schober, J. Yuan and V. K. Bhargava, "A Survey on Non-Orthogonal Multiple Access for 5G Networks: Research Challenges and Future Trends," in IEEE Journal on Selected Areas in Communications, vol. 35, no. 10, pp. 2181-2195, Oct. 2017.

\bibitem{b8} J. N. Laneman, D. N. C. Tse and G. W. Wornell, "Cooperative diversity in wireless networks: Efficient protocols and outage behavior," in IEEE Transactions on Information Theory, vol. 50, no. 12, pp. 3062-3080, Dec. 2004.

\bibitem{b9} M. F. Kader, M. B. Uddin, S. R. Islam, and S. Y. Shin, “Capacity and outage analysis of a dual-hop decode-and-forward relay-aided NOMA scheme,” Digital Signal Processing, vol. 88, pp. 138–148, 2019.

\bibitem{b10} R. Jiao, L. Dai, J. Zhang, R. MacKenzie and M. Hao, "On the Performance of NOMA-Based Cooperative Relaying Systems Over Rician Fading Channels," in IEEE Transactions on Vehicular Technology, vol. 66, no. 12, pp. 11409-11413, Dec. 2017.


\bibitem{b11} T. N. Do, D. B. da Costa, T. Q. Duong and B. An, "Improving the Performance of Cell-Edge Users in MISO-NOMA Systems Using TAS and SWIPT-Based Cooperative Transmissions," in IEEE Transactions on Green Communications and Networking, vol. 2, no. 1, pp. 49-62, March 2018.

\bibitem{b12} N. T. Do, D. Benevides da Costa, T. Q. Duong and B. An, "Transmit antenna selection schemes for MISO-NOMA cooperative downlink transmissions with hybrid SWIPT protocol," 2017 IEEE International Conference on Communications (ICC), Paris, 2017, pp. 1-6.

\bibitem{b13} A. A. Amin and S. Y. Shin, "Investigate the Dominating Factor of Hybrid SWIPT Protocol by Performance Analysis of the Far User of Hybrid SWIPT based CNOMA Downlink Transmission," 2019 International Conference on Electrical, Computer and Communication Engineering (ECCE), Cox'sBazar, Bangladesh, 2019, pp. 1-6.

\bibitem{b14} Y. Zhang, S. Feng and W. Tang, "Performance Analysis and Optimization for Power Beacon-Assisted Wireless Powered Cooperative NOMA Systems," in IEEE Access, vol. 8, pp. 198436-198450, 2020, doi: 10.1109/ACCESS.2020.3034917. 

\bibitem{b15} Do DT, Le CB. Application of NOMA in Wireless System with Wireless Power Transfer Scheme: Outage and Ergodic Capacity Performance Analysis. Sensors (Basel). 2018 Oct 17;18(10):3501. doi: 10.3390/s18103501. PMID: 30336586; PMCID: PMC6210380.

\bibitem{b16} H. Liu, Z. Ding, K. J. Kim, K. S. Kwak and H. V. Poor, "Decode-and-Forward Relaying for Cooperative NOMA Systems With Direct Links," in IEEE Transactions on Wireless Communications, vol. 17, no. 12, pp. 8077-8093, Dec. 2018.

\bibitem{b17} Z. Yang, Z. Ding, Y. Wu and P. Fan, "Novel relay selection strategies for cooperative NOMA", IEEE Trans. Veh. Technol., vol. 66, no. 11, pp. 10114-10123, Nov. 2017.

\bibitem{b18} J. Tang et al., "Energy Efficiency Optimization for NOMA With SWIPT", in IEEE Journal of Selected Topics in Signal Processing, vol. 13, no. 3, pp. 452-466, June 2019, doi: 10.1109/JSTSP.2019.2898114.

\bibitem{b19} X. Lu, P. Wang, D. Niyato, D. I. Kim and Z. Han, "Wireless Networks With RF Energy Harvesting: A Contemporary Survey," in IEEE Communications Surveys and Tutorials, vol. 17, no. 2, pp. 757-789, Second quarter 2015, doi: 10.1109/COMST.2014.2368999.

\bibitem{b20} M. F. Kader, M. B. Uddin, A. Islam, and S. Y. Shin, “Cooperative non‐orthogonal multiple access with SWIPT over Nakagami‐ m fading channels,” Transactions on Emerging Telecommunications Technologies, vol. 30, no. 5, 2019.

\bibitem{b21} Z. Yang, Z. Ding, P. Fan and N. Al-Dhahir, "The Impact of Power Allocation on Cooperative Non-orthogonal Multiple Access Networks With SWIPT," in IEEE Transactions on Wireless Communications, vol. 16, no. 7, pp. 4332-4343, July 2017, doi: 10.1109/TWC.2017.2697380.

\bibitem{b22} I. Abu Mahady, E. Bedeer, S. Ikki and H. Yanikomeroglu, "Sum-Rate Maximization of NOMA Systems Under Imperfect Successive Interference Cancellation," in IEEE Communications Letters, vol. 23, no. 3, pp. 474-477, March 2019, doi: 10.1109/LCOMM.2019.2893195.

\bibitem{b23} G. Li and D. Mishra, "Rate-Constrained Energy Minimization in Hybrid SWIPT for Relay-Assisted NOMA Networks," 2020 IEEE International Conference on Communications Workshops (ICC Workshops), Dublin, Ireland, 2020, pp. 1-6, doi: 10.1109/ICCWorkshops49005.2020.9145250. 

\bibitem{b24} Z. Ding, M. Peng and H. V. Poor, "Cooperative Non-Orthogonal Multiple Access in 5G Systems," in IEEE Communications Letters, vol. 19, no. 8, pp. 1462-1465, Aug. 2015.

\bibitem{b25} M. F. Kader and S. Y. Shin, "Coordinated Direct and Relay Transmission Using Uplink NOMA," in IEEE Wireless Communications Letters, vol. 7, no. 3, pp. 400-403, June 2018.

\bibitem{b26} M. F. Kader, S. Y. Shin and V. C. M. Leung, "Full-Duplex Non-Orthogonal Multiple Access in Cooperative Relay Sharing for 5G Systems," in IEEE Transactions on Vehicular Technology, vol. 67, no. 7, pp. 5831-5840, July 2018, doi: 10.1109/TVT.2018.2799939.

\bibitem{b27} M. B. Shahab and S. Y. Shin, "A Time Sharing Based Approach to Accommodate Similar Gain Users in NOMA for 5G Networks," 2017 IEEE 42nd Conference on Local Computer Networks Workshops (LCN Workshops), Singapore, 2017, pp. 142-147, doi: 10.1109/LCN.Workshops.2017.76.

\bibitem{b28} Y. Huang, M. Liu and Y. Liu, "Energy-Efficient SWIPT in IoT Distributed Antenna Systems," in IEEE Internet of Things Journal, vol. 5, no. 4, pp. 2646-2656, Aug. 2018.

\end{thebibliography}
\end{document}